\documentclass[preprint,aps]{revtex4}
\usepackage{amsmath}
\usepackage{graphicx}
\usepackage{epsfig}

\begin{document}
\baselineskip=24ptplus.5ptminus.2pt \vspace*{0.1 in} \large
\begin{center}
{\bf Extended Optical Model Analyses of \\
Elastic Scattering, Direct Reaction, and Fusion \\
Cross Sections for the $^{9}$Be+$^{208}$Pb System \\
at Near-Coulomb-Barrier Energies}
\end{center}
\normalsize
\vspace{0.5cm}

\begin{center}
W. Y. So\\
{\it Laboratory of Cyclotron Application, \\
Korea Institute of Radiological and Medical Sciences, Seoul 139-706, Korea}\\
\parskip 2ex
S. W. Hong, and B. T. Kim\\
{\it Department of Physics and Institute of Basic Science, \\
Sungkyunkwan University, Suwon 440-746, Korea}\\
\parskip 4.5ex
T. Udagawa \\
{\it Department of Physics, University of Texas, Austin, Texas
78712}
\end{center}
\begin{center}
{\bf Abstract}
\end{center}

\baselineskip=24ptplus.5ptminus.2pt

Based on the extended optical model approach in which the
polarization potential is decomposed into direct reaction (DR) and
fusion parts, simultaneous $\chi^{2}$ analyses are performed for
elastic scattering, DR, and fusion cross section data for the
$^{9}$Be+$^{208}$Pb system at near-Coulomb-barrier energies.
Similar $\chi^{2}$ analyses are also performed by only taking into account
the elastic scattering and fusion data as was previously done by
the present authors, and the results are compared with those of
the full analysis including the DR cross section data as well.
We find that the analyses using only elastic scattering and fusion data  
can produce very consistent and reliable predictions of cross sections
particularly when the DR cross section data are not complete.
Discussions are also given on the results obtained from similar
analyses made earlier for the $^{9}$Be+$^{209}$Bi system.

\vspace{1.5cm} PACS numbers : 24.10.-i,~25.70.Jj

\pagebreak
\section{Introduction}

In our recent study~\cite{so1}, we have carried out simultaneous
$\chi^{2}$ analyses of elastic scattering and fusion cross section
data for the $^{6}$Li+$^{208}$Pb~\cite{kee1,wu1,das1} and
$^{9}$Be+$^{209}$Bi~\cite{sig1,sig2} systems at
near-Coulomb-barrier energies in the framework of an extended
optical model~\cite{uda1,hong,uda2} by introducing two types of
complex polarization potentials: the direct reaction (DR) and
fusion potentials. In such analyses, it is indispensable and essential to
include the experimental data for the total DR cross section
$\sigma_{D}^{exp}$ and the fusion cross section
$\sigma_{F}^{exp}$, in addition to the elastic scattering cross section
$d\sigma_{E}^{exp}/d\Omega$ for the separate determination of the
DR and fusion potentials. However, when the previous study was
made \cite{so1}, reliable data of
$\sigma_{D}^{exp}$ for $^{6}$Li+$^{208}$Pb and
$^{9}$Be+$^{209}$Bi were not available, and thus
the analyses were proceeded in two steps. In the
first step, we carried out $\chi^{2}$ analyses of only the elastic
scattering data by assuming just one simple Woods-Saxon type
complex potential. Using the potential parameters thus fixed, we could
then generate the total reaction cross section $\sigma_{R}$, which we
called the semi-experimental total reaction cross section
$\sigma_{R}^{semi-exp}$. As has been shown in a number of
publications, such $\sigma_{R}^{semi-exp}$ predicted from the optical
potential that fits the elastic scattering data usually reproduces
$\sigma_{R}^{exp}$ very well. 
This is the case for reactions
induced by the proton~\cite{mcc1,eli1}, the deuteron~\cite{auc1},
the $\alpha$-particle~\cite{sin1}, and also heavy-ions~\cite{kol1}.
We then used $\sigma_{R}^{semi-exp}$ to further extract
semi-experimental total DR cross sections $\sigma_{D}^{semi-exp}$
by using the relation
$\sigma_{D}^{semi-exp}=\sigma_{R}^{semi-exp}-\sigma_{F}^{exp}$. 
In the second step, use was made of thus extracted
$\sigma_{D}^{semi-exp}$ in place of the experimental DR cross section
$\sigma_{D}^{exp}$ to carry out simultaneous analyses of
$d\sigma_{E}^{exp}/d\Omega$, $\sigma_{D}^{semi-exp}$, and
$\sigma_{F}^{exp}$ for determining the full extended optical model
potential composed of two polarization potentials.

The DR and fusion potentials thus determined revealed interesting
characteristic features of these potentials. First of all, both
potentials satisfy separately the dispersion relation~\cite{maha}.
Secondly, the fusion potential is found to exhibit a threshold
anomaly~\cite{maha,naga}, as was observed for tightly bound
projectiles~\cite{bae1,lil1,ful1}, but the DR potential does not
show a pronounced threshold anomaly. Thirdly at the strong
absorption radius, the magnitudes of the fusion potential were
found to be much smaller than those of the DR potential. As a
consequence, the resulting total polarization potential dominated
by the DR potential becomes rather smooth as a function of the
incident energy. This has solved a long standing puzzle why the
threshold anomaly has not been seen in the polarization potentials
determined for the systems involving a loosely bound projectile such
as $^{6}$Li and $^{9}$Be~\cite{kee1,sig1}.

The extracted DR potentials have provided us with a unique
opportunity to study the effects of breakup (DR) on fusion by
comparing $\sigma_{F}$ calculated from either including or
neglecting the real and the imaginary part of the DR potential. Such
studies were made in Ref.~\cite{so1}, which showed that in the
sub-barrier region, breakup is not the main reason for the
subbarrier enhancement of $\sigma_{F}^{exp}$ and that the
mechanism that governs the enhancement is neutron flow as
originally suggested by Stelson {\it et al.}~\cite{stel}. In our
approach, this effect is phenomenologically implemented in the
imaginary part of the DR potential. On the other hand, in the
above barrier region, the breakup suppresses $\sigma_{F}^{exp}$ and
the observed suppression factors for $^{6}$Li and $^{9}$Be were
fairly well accounted for in terms of the breakup \cite{so1}.

After completing our work of Ref.~\cite{so1} for $^9$Be +
$^{209}$Bi system, elastic scattering data for $^9$Be + $^{208}$Pb, 
a system similar to $^9$Be + $^{209}$Bi, have become available~\cite{woo1}. 
Thus, for $^9$Be + $^{208}$Pb system we now have data available for the elastic
scattering~\cite{woo1} and fusion~\cite{das2} cross sections as well as
the sum of cross sections of breakup, transfer, and
incomplete fusion~\cite{woo2}. For a loosely bound projectile like
$^{9}$Be, we may assume the summed cross section thus observed as
the total DR cross sections. This has provided us with an
opportunity to carry out $\chi^{2}$ analyses taking into account
all three sets of experimental data, i.e., the elastic scattering,
DR, and fusion data. We can then compare these $\chi^{2}$ analysis results 
with those obtained by considering only two sets of data, i.e., the elastic
scattering and fusion data without the experimental DR data. 
We shall henceforth call the case
where all the three data sets are included in the $\chi^{2}$ analyses the EDF
(elastic, DR, and fusion) approach, and the case where only two
data sets are considered the EF (elastic scattering and fusion) approach. 
Note that in the EF approach we do, however, include 
$\sigma_{D}^{semi-exp}$, which is
essential to fix the DR potential parameters.
The aim of the present study is to make a comparison
between these two approaches and study the validity of the EF
method used in our previous work. 
By extending our EF method proposed in the previous work 
on  $^9$Be + $^{209}$Bi to $^9$Be + $^{208}$Pb system, 
where we have DR data as well as elastic and fusion data,
we shall show that the EF approach gives us very reliable predictions 
of cross sections.

In Sec. II, we first generate $\sigma_{D}^{semi-exp}$ for the EF
approach case by following the method described in
Ref.~\cite{so1}. Two types of $\chi^{2}$ analyses (EDF and EF) are then
carried out in Sec. III and the results are compared and discussed in Sec. IV.
Sec. V concludes the paper.

\section{Extracting semi-experimental DR cross section}

Our method of generating $\sigma_{D}^{semi-exp}$ resorts to the
empirical fact~\cite{brog} that the total reaction cross section
calculated from the optical model fit to the available elastic
scattering cross section data, $d\sigma^{exp}_{E}/d\Omega$,
usually agrees well with the experimental $\sigma_{R}$, in spite
of the well known ambiguities in the optical potential. Let us
call the total reaction cross section
thus generated the semi-experimental reaction
cross section $\sigma_{R}^{semi-exp}$. Then,
$\sigma_{D}^{semi-exp}$ is generated by
\begin{equation}
\sigma^{semi-exp}_{D} = \sigma^{semi-exp}_{R} - \sigma^{exp}_{F}.
\end{equation}
This approach seems to work even for loosely bound projectiles, as
demonstrated recently by Kolata {\it et al.}~\cite{kol1} for the
$^{6}$He+$^{209}$Bi system.

Following Ref.~\cite{so1}, we first carry out rather simple
optical model $\chi^{2}$ analyses of elastic scattering data
solely for the purpose of deducing $\sigma_{R}^{semi-exp}$. For
these preliminary analyses, we assume the optical potential to be
a simple sum of two volume-type potentials $V_{0}(r)$ and
$U_{1}(r,E)$, where $V_{0}(r)$ is the real, energy independent
bare potential and $U_{1}(r,E)$ is a complex potential with
common geometrical parameters for both real and imaginary parts.
The elastic scattering data are then fitted with a fixed radius
parameter $r_{1}$ for $U_{1}(r,E)$ and with three other
parameters adjustable; the real and the imaginary strengths
$V_{1}$ and $W_{1}$ and the diffuseness parameter $a_{1}$. The
$\chi^{2}$ fitting is done for three choices of the radius
parameter; $r_{1}$=1.3, 1.4, and 1.5 fm. These different choices
of the $r_{1}$-value are made in order to examine the dependence
of the resulting $\sigma_{R}^{semi-exp}$ on the value of $r_1$.

\begin{table}
\caption{Measured and extracted fusion, DR, and total reaction
cross sections for the $^{9}$Be+$^{208}$Pb system.
$\sigma_{F}^{exp}$ and $\sigma_{D}^{exp}$ are from 
Refs.~\cite{das2} and \cite{woo2}, respectively.
$\sigma_{R}^{exp}$ is the sum of $\sigma_{F}^{exp}$ and 
$\sigma_{D}^{exp}$. $\sigma_{R}^{semi-exp}$ is 
extracted from the elastic scattering data~\cite{woo1} as explained in the text.
$\sigma_{D}^{semi-exp}$ is then obtained by using Eq.~(1).
} \vspace{2ex}
\label{table1}
\begin{ruledtabular}
\begin{tabular}{ccccccc}
$E_{lab} $ &$E_{c.m.}$ & $\sigma^{exp} _{F}$ & $\sigma^{exp} _{D}$ & $\sigma^{semi-exp}_{D}$ & $\sigma^{exp}_{R}$ & $\sigma^{semi-exp}_{R}$ \\
(MeV) & (MeV) & (mb) & (mb) & (mb) & (mb) & (mb) \\ \hline
38 & 36.4 & 10 & 109 & 71 & 119 & 81 \\
40 & 38.3 & 58 & 180 & 198 & 238 & 256 \\
42 & 40.3 & 145 & 267 & 320 & 412 & 465 \\
44 & 42.2 & 248 & 300 & 368 & 548 & 616 \\
46 & 44.1 & 355 & 300 & 423 & 655 & 778 \\
48 & 46.0 & 458 & 360 & 541 & 818 & 999 \\
50 & 47.9 & 580 & 410 & 597 & 990 & 1177 \\
\end{tabular}
\end{ruledtabular}
\end{table}

As observed in Ref.~\cite{so1}, the values of
$\sigma_{R}^{semi-exp}$ thus extracted for three different
$r_{1}$-values agree with the average 
within 1\%, implying that
$\sigma_{R}^{semi-exp}$ is determined without much ambiguity. We
then identified the average as the final value of
$\sigma_{R}^{semi-exp}$. Using thus determined
$\sigma_{R}^{semi-exp}$, we generated $\sigma_{D}^{semi-exp}$ by
employing Eq.~(1). The resultant values of $\sigma_{R}^{semi-exp}$
and $\sigma_{D}^{semi-exp}$ are presented in Table I, together with
$\sigma_{F}^{exp}$~\cite{das2}, 
$\sigma_{D}^{exp}$~\cite{woo2}, 
and
$\sigma_{R}^{exp}$. As seen from Table I, the values of
$\sigma_{D}^{semi-exp}$ and $\sigma_{R}^{semi-exp}$ are
systematically larger than the corresponding experimental values,
except for the lowest energy of $E_{cm}$=36.4~MeV. The reason why
$\sigma_{D}^{semi-exp}$ is larger than $\sigma_{D}^{exp}$ except
for the lowest energy may be ascribed
to the fact that $\sigma_{D}^{exp}$ includes contributions from
only breakup, transfer, and incomplete-fusion events~\cite{woo2}, but not from
inelastic scattering and other simple quasi-elastic processes such
as pickups.
The difference between $\sigma_{R}^{exp}$ and $\sigma_{R}^{semi-exp}$
becomes larger with energy. It implies that there are more open, but not
identified, DR channels as the incident energy increases.

It is worth remarking at this point that there is a reason to
question the accuracy of the extracted value of $\sigma_{D}^{semi-exp}$ at
$E_{cm}$=36.4~MeV. The experimental value of the ratio
$P_{E}$ of the elastic scattering to the Rutherford cross section
at the forward angles are systematically larger than unity~\cite{woo1}
at this energy.
The average value of $P_{E}$ at small angles is about 1.033. 
This suggests that there may be a
problem in the overall normalization constant in the measured
data. In fact, it is indicated~\cite{woo1} that there
are experimental uncertainties of a few percents in the absolute
normalization. Even just a few percent uncertainty in the normalization
is critical, particularly at low energies in extracting
$\sigma_{D}^{semi-exp}$. In order to confirm this, we
have reanalyzed the elastic scattering data by reducing 
the cross section by a factor of 1.033 
so that the values of $P_{E}$ at forward angles become
around unity. A new value of $\sigma_{R}^{semi-exp}$ thus extracted
turns out to be 122~mb, which in turn gives us
$\sigma_{D}^{semi-exp}$=112~mb. This value is significantly greater than
71 mb given in Table I and is also larger than the
experimental value of $\sigma_{D}^{semi-exp} =$ 109~mb. 
It is thus very plausible that the true values of
$\sigma_{D}^{semi-exp}$ and $\sigma_{R}^{semi-exp}$ at this energy 
could be larger than $\sigma_{D}^{exp}$ and $\sigma_{R}^{exp}$, respectively.
However, in the present $\chi^{2}$ analyses, use
is made of the $\sigma_{D}^{semi-exp}$ values as listed in Table I.

\section{Simultaneous $\chi^{2}$ Analyses}

Simultaneous $\chi^{2}-$analyses are then performed for two cases of data
sets; ($d\sigma^{exp}_{E}/d\Omega$, ~$\sigma_{D}^{exp}$,
~$\sigma^{exp}_{F}$) and ($d\sigma^{exp}_{E}/d\Omega$,
~$\sigma_{D}^{semi-exp}$, ~$\sigma^{exp}_{F}$) where 
$d\sigma^{exp}_{E}/d\Omega$, $\sigma_{D}^{exp}$, and
$\sigma^{exp}_{F}$ are from the literatures~\cite{woo1,das2,woo2}.
As mentioned in the Introduction, 
the former (latter) case with $\sigma_{D}^{exp}$ ($\sigma_{D}^{semi-exp}$)
is called the EDF (EF) analysis. In calculating the $\chi^{2}$
value, we simply assume 1\% errors for all the experimental data.
The 1\% error is roughly the average of errors in the measured
elastic scattering cross sections, but it is much smaller than the
errors in the DR ($\sim$5\%) and fusion ($\sim$10\%) cross
sections. The choice of the 1\% error for DR and fusion cross
sections is thus equivalent to increasing the weight for the DR
and fusion cross sections in evaluating the $\chi^{2}$-values by
factors of 25 and 100, respectively. Such a choice of errors may
be reasonable, since we have only one datum point for each of
these cross sections, while there are more than 50 data points for
the elastic scattering cross sections.

\subsection{Necessary Formulae}

The optical potential $U(r,E)$ we use in the present work has the
following form;
\begin{equation}
U(r;E) = V_{C}(r)-[V_{0}(r)+U_{F}(r;E)+U_{D}(r;E)],
\end{equation}
where $V_{C}(r)$ is the usual Coulomb potential with $r_{C}$=1.25
fm and $V_{0}(r)$ is the bare (Hartree-Fock) nuclear potential.
$U_{F}(r;E)$ and $U_{D}(r;E)$ are, respectively, fusion and DR
parts of the polarization potential~\cite{love} originating
from couplings to the respective reaction channels.
Both $U_{F}(r;E)$ and $U_{D}(r;E)$ are complex and their forms are
assumed to be of volume-type and
surface-derivative-type~\cite{hong,kim1}, respectively.
$V_{0}(r)$, $U_{F}(r;E)$, and $U_{D}(r;E)$ are explicitly given by
\begin{equation}
V_{0}(r)=V_{0}f(X_{0}),
\end{equation}
\begin{equation}
U_{F}(r;E) = (V_{F}(E)+iW_{F}(E))f(X_{F}),
\end{equation}
and
\begin{equation}
U_{D}(r;E) = (V_{D}(E)+iW_{D}(E))4a_{D}\frac{df(X_{D})}{dR_{D}},
\vspace{2ex}
\end{equation}
where $f(X_{i})=[1+\mbox{exp}(X_{i})]^{-1}$ with
$X_{i}=(r-R_{i})/a_{i}$ $({\it i}=0, \; D\; \mbox{and} \; F)$ is
the usual Woods-Saxon function, while $V_{F}(E)$, $V_{D}(E)$,
$W_{F}(E)$, and $W_{D}(E)$ are the energy-dependent strength
parameters. We assume the geometrical parameters of the real and
imaginary potentials are the same, and thus the strength
parameters $V_{i}(E)$ and $W_{i}(E)$ ($i=F$ or $D$) are related
through a dispersion relation~\cite{maha},
\begin {equation}
V_{i}(E)=V_{i}(E_{s}) + \frac {E-E_{s}}{\pi } \mbox{P}
\int_{0}^{\infty} dE' \frac {W_{i}(E')}{(E'-E_{s})(E'-E)},
\vspace{2ex}
\end {equation}
where P stands for the principal value and $V_{i}(E_{s})$ is the
value of $V_{i}(E)$ at a reference energy $E=E_{s}$. Later, we
will use Eq.~(6) to generate the final real strength parameters
$V_{F}(E)$ and $V_{D}(E)$, after $W_{F}(E)$ and $W_{D}(E)$ are
fixed from $\chi^{2}$ analyses. Note that the breakup cross
section may include contributions from both Coulomb and nuclear
interactions, which implies that the direct reaction potential
includes effects coming from not only the nuclear interaction, but
also the Coulomb interaction.

$V_{0}(r)$ in Eq.~(3) may also have an energy-dependence coming
from the nonlocality due to the knockon-exchange contribution. We
ignore such effects as they are expected to be small for heavy-ion
scattering~\cite{sat1}, and employ the real potential parameters
used in Ref.~\cite{wo77} assuming that all the unusual features of
the potential may be put into the polarization parts, particularly in
the DR part. The parameters used for $V_{0}(r)$ are $V_{0}$=18.36~MeV,
$r_{0}$=1.22~fm, and $a_{0}$=0.57~fm ~\cite{wo77}. Note that this
potential is shallow, which is often required in fitting elastic
scattering data of such projectiles as $^{6}$Li and
$^{9}$Be~\cite{sat2}.

In performing the optical model calculation, one can evaluate
$\sigma_{F}$ and $\sigma_{D}$ by using the following
expression~\cite{uda1,hong,uda2,huss}
\begin {equation}
\sigma_{i} = \frac {2}{\hbar v} <\chi \raisebox{1ex}{(+)}|W_{\it
i}(r)|\chi \raisebox{1ex}{(+)}> \hspace{.5in} (i=F\;\mbox{or}\;D),
\end{equation}
where $\chi^{(+)}$ is the usual distorted wave function that
satisfies the Schr\"{o}dinger equation with the full optical model
potential $U(r,E)$ in Eq.~(2). $\sigma_{F}$ and $\sigma_{D}$ are
thus calculated within the same framework as
$d\sigma_{E}/d\Omega$ is calculated. Such a unified description
enables us to treat different types of reactions on the
same footing.

\subsection{Threshold Energies of Subbarrier Fusion and DR}

As in Ref.~\cite{so1}, we also utilize as an important ingredient
the so-called threshold energies $E_{0,F}$ and $E_{0,D}$ of
subbarrier fusion and DR, respectively, which are defined as zero
intercepts of the linear representation of the quantities
$S_{i}(E)$, defined by
\begin{equation}
S_{i} \equiv \sqrt{E \sigma_{i}} \approx \alpha_{i} (E-E_{0,i})
\;\;\; (i=D \; \mbox{or} \; F),
\end{equation}
where $\alpha_{i}$ is a constant. $S_{i}$ with $i=F$, i.e.,
$S_{F}$ is the quantity introduced originally by Stelson {\it et
al.}~\cite{stel}, who showed that in the subbarrier region $S_{F}$
from the measured $\sigma_{F}$ can be represented very well by a
linear function of $E$ (linear systematics) as in Eq.~(8). In
Ref.~\cite{kim1}, we extended the linear systematics to DR cross
sections. In fact the DR data are also well represented by a
linear function.

In Fig.~1(a), we present the experimental $S_{F}(E)$ and $S_{D}(E)$.
From the zeros of $S_{i}(E)$, one can deduce
$E_{0,D}^{exp}$=30.0~MeV and $E_{0,F}^{exp}=$35.0~MeV. For both
$i=F$ and $D$, the observed $S_{i}$ are very well approximated by
straight lines in the subbarrier region and thus $E_{0,i}$ can be
extracted without much ambiguity. Another determination of
$E_{0,D}^{exp}$ can be made by using the semi-experimental DR
cross section instead of the experimental DR cross section, 
as shown in Fig.~1(b). The resultant value, which
we shall denote by $E_{0,D}^{semi-exp}$ is found to be
$E_{0,D}^{semi-exp}=32.5$ MeV, close to $E_{0,D}^{exp}$.

$E_{0,i}$ may then be used as the energy where the imaginary
potential $W_{i}(E)$ becomes zero, i.e.,
$W_{i}(E_{0,i})=0$~\cite{kim1,kim2}. This procedure will be used
later in obtaining a mathematical expression for $W_{i}(E)$.

\subsection{$\chi^{2}$ Analyses}

All the $\chi^{2}$ analyses performed in the present work are
carried out by using $V_{0}(r)$ as given in Subsec.~III~A and by
using the fixed geometrical parameters for the polarization
potentials, $r_{F}$=1.40~fm, $a_{F}$=0.30~fm, $r_{D}$=1.50~fm, and
$a_{D}$=0.70~fm, which are close to the values used in our
previous study~\cite{so1}. 
Small changes of these values from the ones used in
Ref.~\cite{so1} are made in order to improve the
$\chi^{2}$-fitting.

As in Ref.~\cite{so1}, the $\chi^{2}$ analyses are done in two
steps; in the first step, all 4 strength parameters, $V_{D}(E)$,
$W_{D}(E)$, $V_{F}(E)$ and $W_{F}(E)$ are varied. In this step,
we have been able to fix the strength parameters of the DR potential, 
$V_{D}(E)$ and $W_{D}(E)$, fairly well in the sense that the extracted 
$V_{D}(E)$ and $W_{D}(E)$ turn out to be smooth as functions of $E$. 
This is particularly the case for the imaginary strength $W_{D}(E)$.
The values of $V_{D}(E)$ and $W_{D}(E)$ are
presented in Figs.~2 and 3 by open circles for the EDF and EF
cases, respectively. It is remarkable that the resultant
$W_{D}(E)$ can be fairly well represented by the following
function of $E(=E_{c.m.})$ (in units of MeV)
\begin{equation} W_{D}(E) \; = \; \left \{
\begin{array}{lll}
0 &\;\; \mbox{for $E\leq E_{0, D}^{exp}=$30.0} \\
0.037(E-30.0) &\;\; \mbox{for 30.0$<E\leq$39.0} \\
0.33 &\;\; \mbox{for 39.0$< E$} \\
\end{array}
\right. \vspace{2ex}
\end{equation}
in the EDF case and
\begin{equation}
W_{D}(E) \; = \; \left \{ \begin{array}{lll}
0 &\;\; \mbox{for $E \leq E_{0, D}^{semi-exp}=$32.5} \\
0.052(E-32.5) &\;\; \mbox{for 32.5$<E\leq$40.0} \\
0.39 &\;\; \mbox{for 40.0$< E$} \\
\end{array}
\right. \vspace{2ex}
\end{equation}
in the EF case. Note that the threshold energies where $W_{D}(E)$
becomes zero are set equal to $E_{0,D}^{exp}$ and
$E_{0,D}^{semi-exp}$ as determined in the previous subsection 
and are indicated by the open half circles sitting on the axis of $E_{c.m.}$ 
in Figs. 2 and 3. The dotted lines in the lower panels of Figs.~2 and 3 
represent Eqs.~(9) and (10), respectively. The dotted curves 
in the upper panels of Figs.~2 and 3 denote $V_{D}$ 
as predicted by the dispersion
relation Eq.~(6), with $W_{D}(E)$ given by Eqs.~(9) and (10),
respectively. As seen, the dotted curves reproduce the open circles
fairly well, indicating that $V_{D}(E)$ and $W_{D}(E)$ extracted
by the $\chi^{2}$ analyses satisfy the dispersion relation.

In this first step of $\chi^{2}$ fitting, however, 
$V_{F}(E)$ and $W_{F}(E)$ are not well fixed in the sense
that the extracted values fluctuate considerably as functions of
$E$. This is understandable from the expectation that the elastic
scattering data can probe most accurately the optical potential
in the peripheral region, which is nothing but the region
characterized by the DR potential with $r_D = 1.5$ fm. 
The part of the nuclear potential responsible for fusion with $r_F = 1.4$ fm
is thus difficult to pin down in this first step. 

In order to obtain more reliable information on $V_{F}$ and
$W_{F}$, we have thus performed the second step of the $\chi^{2}$
analysis. This time, instead of doing a 4-parameter search 
we use $V_{D}$ and $W_{D}$ determined by the first step of $\chi^{2}$ fitting.
But, rather than using $V_{D}$ and $W_{D}$ exactly as determined
by the $\chi^2$ fitting, we use $W_{D}(E)$ given by Eqs.~(9) and (10) and
$V_{D}(E)$ given by the dispersion relation. We then have performed
2-parameter $\chi^{2}$ analyses, treating only $V_{F}(E)$ and
$W_{F}(E)$ as adjustable parameters. The values thus determined
are presented in Figs.~2 and 3 by solid circles. As seen, both
$V_{F}(E)$ and $W_{F}(E)$ are determined to be fairly smooth
functions of $E$. The extracted $W_{F}(E)$ may be represented by
\begin{equation}
W_{F}(E) \; = \; \left \{ \begin{array}{lll}
0 &\;\; \mbox{for $E\leq E_{0, F}^{exp}=$35.0} \\
0.879(E-35.0) &\;\; \mbox{for 35.0$<E\leq$38.3} \\
2.90 &\;\; \mbox{for 38.3$< E$} \\
\end{array}
\right. \vspace{2ex}
\end{equation}
in the EDF case and
\begin{equation}
W_{F}(E) \; = \; \left \{ \begin{array}{lll}
0 &\;\; \mbox{for $E\leq E_{0, F}^{exp}=$35.0} \\
0.771(E-35.0) &\;\; \mbox{for 35.0$<E\leq$38.5} \\
2.70 &\;\; \mbox{for 38.5$< E$} \\
\end{array}
\right. \vspace{2ex}
\end{equation}
in the EF case, respectively. As is done for $W_{D}(E)$, the
threshold energy where $W_{F}(E)$ becomes zero is set equal to
$E_{0,F}^{exp}$ and is indicated in Figs.~2 and 3 
by the solid half circle on the axis of $E_{c.m.}$. 
As seen, the $W_{F}(E)$ values determined by the
second $\chi^{2}$ analyses are fairly well represented by the
functions given by Eqs.(11) and (12). Note that the
energy variations of $W_{F}(E)$ and $V_{F}(E)$ are quite
rapid compared to those of $W_{D}(E)$ and $V_{D}(E)$, and are
similar to those observed in tightly bound
projectiles~\cite{bae1,lil1,ful1}.

Using $W_{F}(E)$ given by Eqs.~(11) and (12), one can generate
$V_{F}(E)$ from the dispersion relation. The results are shown by
the solid curves in the upper panels of Figs.~2 and 3, which 
well reproduce the solid circles extracted from the $\chi^{2}$-fitting.
This means that the fusion potential determined from the present
analysis satisfies the dispersion relation.

\subsection{Final Calculated Cross Sections in Comparison with
the Data}

Using $W_{D}(E)$ given by Eqs.~(9) and (10) and $W_{F}(E)$ given
by Eqs.~(11) and (12) together with $V_{D}(E)$ and $V_{F}(E)$
generated by the dispersion relation Eq.~(6), we have performed the final
calculations of the elastic, DR and fusion cross sections. 
Thus, instead of using the potential parameters just as extracted by
the $\chi^2$-analyses we have used these dispersive potentials
for the final calculations.
The results are presented in Figs.~4 and 5 in comparison with the
experimental data. All the data are well reproduced by the
calculations, though there are subtle differences between the fits
obtained by two types of the analyses as will be discussed in
detail in Subsec.~IV~B.

\section{Discussions}

\subsection{Fusion and DR Potentials}

The characteristic features of the polarization potentials
determined in the present $\chi^{2}$ analyses are very much
similar to those obtained in our previous analyses~\cite{so1}. 
The real and imaginary parts of both fusion
and DR potentials satisfy well the dispersion relation, and the
fusion potential displays the threshold anomaly. As already
presented in Figs. 2 and 3,
these features are seen in the strength parameters,
$V_{F}(E)$, $W_{F}(E)$, $V_{D}(E)$ and $W_{D}(E)$. 

Another important feature of the extracted potentials is that 
at the strong absorption radius of $R_{sa}$=12.3~fm 
both the real and imaginary parts of the DR potential are 
considerably greater than those of the fusion potential, 
although the strength parameters $V_{D}(E)$ and $W_{D}(E)$ are 
smaller than $V_{F}(E)$ and $W_{F}(E)$. 
Thus, 
the energy dependence of the net polarization potential
(sum of the fusion and DR potentials) at $R_{sa}$ becomes
dominated by the DR potential which has a relatively smooth energy
dependency. Consequently, the net potential does
not show such a threshold anomaly as seen in the net potential for
systems with tightly bound projectiles~\cite{bae1,lil1,ful1}.
However, after separating the polarization potential into DR and fusion parts, 
we clearly observe the characteristic threshold anomaly in the fusion potential.

\subsection{Comparison of EDF and EF Cross Sections}

Both EDF and EF approaches reproduce 
the experimental elastic scattering cross sections very well as shown in Fig.~4. 
The calculated cross sections shown in the left and right panels of Fig.~4
agree well with each other. It may then be
naturally expected that the resultant total reaction cross sections also 
agree with each other. This is indeed the case; 
the values of the calculated total reaction cross
sections from the EDF approach are approximately
equal to those from the EF approach, 
as shown by the dashed curves in Figs.~5(a) and (b).

Since $\sigma_{R}^{semi-exp}$ is extracted from
the fit to the elastic scattering data, 
our final calculation results using the dispersive potential 
naturally reproduce $\sigma_{R}^{semi-exp}$ as shown 
by the dotted curves in Fig. 5(b). 
In the EF case, the calculations  also
reproduce both DR and fusion cross sections as well. 
This is, however, not the case for the EDF approach; 
calculations using the dispersive potential somewhat 
overestimates the experimental data of all three cross sections 
as Fig. 5(a) shows.

It may thus be concluded that the overall fit to the data obtained
in the EF case is better than the EDF case and that the main
source of problems in getting a good overall fit in the EDF case 
comes from inconsistency between the elastic scattering~\cite{woo1}
and the DR reaction~\cite{woo2} data; 
the elastic scattering data require more absorption (larger total
reaction cross section) than what the measured total 
absorption (reaction) cross sections tell us. In view of this, it is
important that measurements be made of inelastic
scattering and some other quasi-elastic reactions which are not
taken into account in the total DR cross section used in the
present analyses.

\subsection{Effects of Breakup on Fusion}

We now turn to the effect of breakup on the fusion cross section.
As has been argued, there are two competing physical effects of
breakup on the fusion cross section, $\sigma_{F}$. The first is
the lowering of the fusion barrier, which tends to enhance
$\sigma_{F}$. The other is the removal of flux from the elastic
into the breakup channel, which suppresses $\sigma_{F}$. Since
the breakup channel dominates DR, these two competing breakup
effects may be represented by the real ($V_{D}(r;E)$) and
the imaginary ($W_{D}(r;E)$) parts of the DR potential;
$V_{D}(r,E)$ can describe precisely the effect of lowering the
barrier, while $W_{D}(r,E)$ the removal of the flux from the
elastic channel.

To see the effects quantitatively, we have introduced in
Ref.~\cite{so1} the following suppression factor $R^{th}$,
\begin{equation}
R^{th}=\sigma_{F}/\sigma_{F}(V_{D}=W_{D}=0),
\end{equation}
where $\sigma_{F}(V_{D}=W_{D}=0)$ is $\sigma_{F}$ obtained by
setting $V_{D}=W_{D}=0$, i.e., neglecting both barrier-lowering
and flux loss effects, while $\sigma_{F}$ is our final calculated
cross section that includes both $V_{D}$ and $W_{D}$. In the
above-barrier region, $R^{th}$ becomes almost constant and here we
present just the average of the $R^{th}$-values at three highest
energies considered in the present study. The values are 0.87 and
0.82 for the EDF and EF cases, respectively. 
Setting $V_D = 0$ reduces $\sigma_F$, while setting $W_D=0$ increases
$\sigma_F$. Thus, the fact that the
$R^{th}$-values are smaller than unity indicates that the flux
loss effect surpasses the barrier-lowering effect in the above
barrier region. The theoretical values may be
compared with the experimental values of $R^{exp}$=0.79, where
$R^{exp}$ is defined as
\begin{equation}
R^{exp}=\sigma^{exp}_{F}/\sigma_{F}(V_{D}=W_{D}=0).
\end{equation}
by using $\sigma_{F}(V_{D}=W_{D}=0)$ fixed from the EF case.
$R^{exp}$ in the EDF case is 0.77, quite close to $R^{exp}$ in the EF case.

Note that the theoretical suppression factor $R^{th}$=0.82 in the case of EF 
agrees very well with the experimental value of $R^{exp}$=0.79. 
It is natural because the calculated $\sigma_{F}$ agrees with
$\sigma_{F}^{exp}$ in the EF case as shown in Fig. 5(b). 
Similarly, the difference between $R^{th}$=0.87 and
$R^{exp}$=0.77 in the EDF case originates from the discrepancy between
$\sigma_{F}^{exp}$ and the calculated $\sigma_{F}$, seen in Fig.~5(a). 
In either case, both
$R^{th}$ and $R^{exp}$ are consistently and considerably smaller
than unity, implying that the observed suppression of $\sigma_{F}$
can be ascribed to the flux loss in the elastic channel to breakup. 
A similar result was also obtained
in Ref.~\cite{so1}.

Although breakup (or DR) is the dominant factor in the
suppression of $\sigma_F$ in the above barrier region, this is
not the case in the sub-barrier region, where the neutron flow
affects fusion dominantly~\cite{stel}, generally
enhancing the sub-barrier fusion. In Ref.~\cite{so1}, it was
proposed that a good measure for the sub-barrier
fusion enhancement is the quantity $\Delta$ defined as
\begin{equation}
\Delta = V_{B} - E_{0,F},
\end{equation}
where $V_{B}$ is the Coulomb-barrier height and $E_{0,F}$ is the
sub-barrier threshold energy discussed in Subsec.~II~B. ~In
Ref.~\cite{so1}, it is demonstrated that $\Delta$ is very well
proportional to the neutron transfer $Q$-value.

\subsection{Comments on the Analyses of
the $^{9}$Be+$^{209}$Bi System Reported in Ref.~[1]}

In Ref.~\cite{so1}, we presented our analyses on the
$^{9}$Be+$^{209}$Bi system using only the elastic scattering~\cite{sig1} 
and fusion cross section data~\cite{sig2} (the EF type analysis).
Since the target nucleus $^{209}$Bi differ from $^{208}$Pb only by
one proton, it is naturally expected that the experimental cross
sections for the two systems should be very similar. 
This is indeed the case for the elastic scattering cross sections; 
no noticeable difference can be found in the data measured for the Pb
target~\cite{woo1} and Bi target~\cite{sig1}. In contrast
to this, the values of the fusion cross section for the Bi target
we used from Ref.~\cite{sig2} at the time of our analyses \cite{so1} are
significantly larger than those for the Pb target reported in
Ref.~\cite{das1}. Recently, however, the fusion cross sections 
for the Bi target are revised~\cite{sig3}, and
the revised values are now very much the same as those of the
Pb target.

Due to this change in the experimental values of 
$\sigma_{F}^{exp}$ for $^{209}$Bi, we have repeated 
our previous analyses for $^9$Be + $^{209}$Bi system, obtaining
now essentially the same results as in the present work for
$^9$Be + $^{208}$Pb. Therefore, we take this opportunity to revise our
previous values of the suppression factor $R$; the new theoretical
value obtained with the revised data is $R^{th}$=0.81, which can
be compared with the new experimental value of $R^{exp}$=0.79. The
corresponding values reported previously in Ref.~\cite{so1} were
$R^{th}$=0.89 and $R^{exp}$=0.92.

\section{Conclusions}

In summary, we have carried out simultaneous $\chi^{2}$ analyses
of elastic scattering, DR (breakup plus incomplete fusion), and
fusion cross sections for the $^{9}$Be+$^{208}$Pb system at
near-Coulomb-barrier energies within the framework of an extended
optical model that introduces the DR and fusion potentials. Two
types of analyses are made; one using the experimental DR cross
section $\sigma_{D}^{exp}$ (EDF case), and the other using the
semi-experimental DR cross section $\sigma_{D}^{semi-exp}$ (EF
case), together with the measured elastic scattering and fusion
cross sections for both cases. In the second type of the
analyses, $\sigma_{D}^{semi-exp}$ is first extracted from simple
optical model fits to the elastic scattering data only. The
extracted $\sigma_{D}^{semi-exp}$ are found to be significantly
larger than $\sigma_{D}^{exp}$.
In spite of this difference between $\sigma_{D}^{exp}$ and
$\sigma_{D}^{semi-exp}$, the resultant DR and fusion potentials
show common features that they satisfy fairly well the
dispersion relation~\cite{maha} and the fusion potentials show the
threshold anomaly as seen in the potentials for systems with
tightly bound projectiles~\cite{naga,bae1,lil1,ful1}. 

For both EDF and EF cases 
the elastic scattering cross sections are equally well reproduced.
However,  the calculated DR, fusion, and total
reaction cross sections fit the corresponding experimental data
well in the EF case, but not in the EDF case. In the latter case, the
calculations overestimate significantly the experimental DR,
fusion, and total reaction cross sections. 
This is because there are some reaction channels that are not
taken into consideration in the present experimental DR data.

Thus as far as we don't have comprehensive 
$\sigma_{D}^{exp}$ available, the EF analysis gives us better overall results
than the EDF analysis. We believe that if the cross section
of inelastic scattering and some other missing reactions that are not
taken into account in the present data of
$\sigma_{D}^{exp}$~\cite{woo2} are measured and used in the analyses, 
both types of analyses will lead to equally good fit to the data. It is
thus highly desirable that such DR data will be measured in near
future in order to test our expectation and thus to justify the
validity of the EF method proposed in Ref.~\cite{so1}.

The authors sincerely thank Drs. Woolliscroft and Dasgupta for
their kindly sending the numerical values of the data they took.
SWH thank TRIUMF for the hospitality where part of the work is done.
The work is supported by the Basic Research Program of the KOSEF,
Korea (Grant No. R05-2003-000-12088-0).

\begin{figure}
\begin{center}
\includegraphics[width=0.75\linewidth] {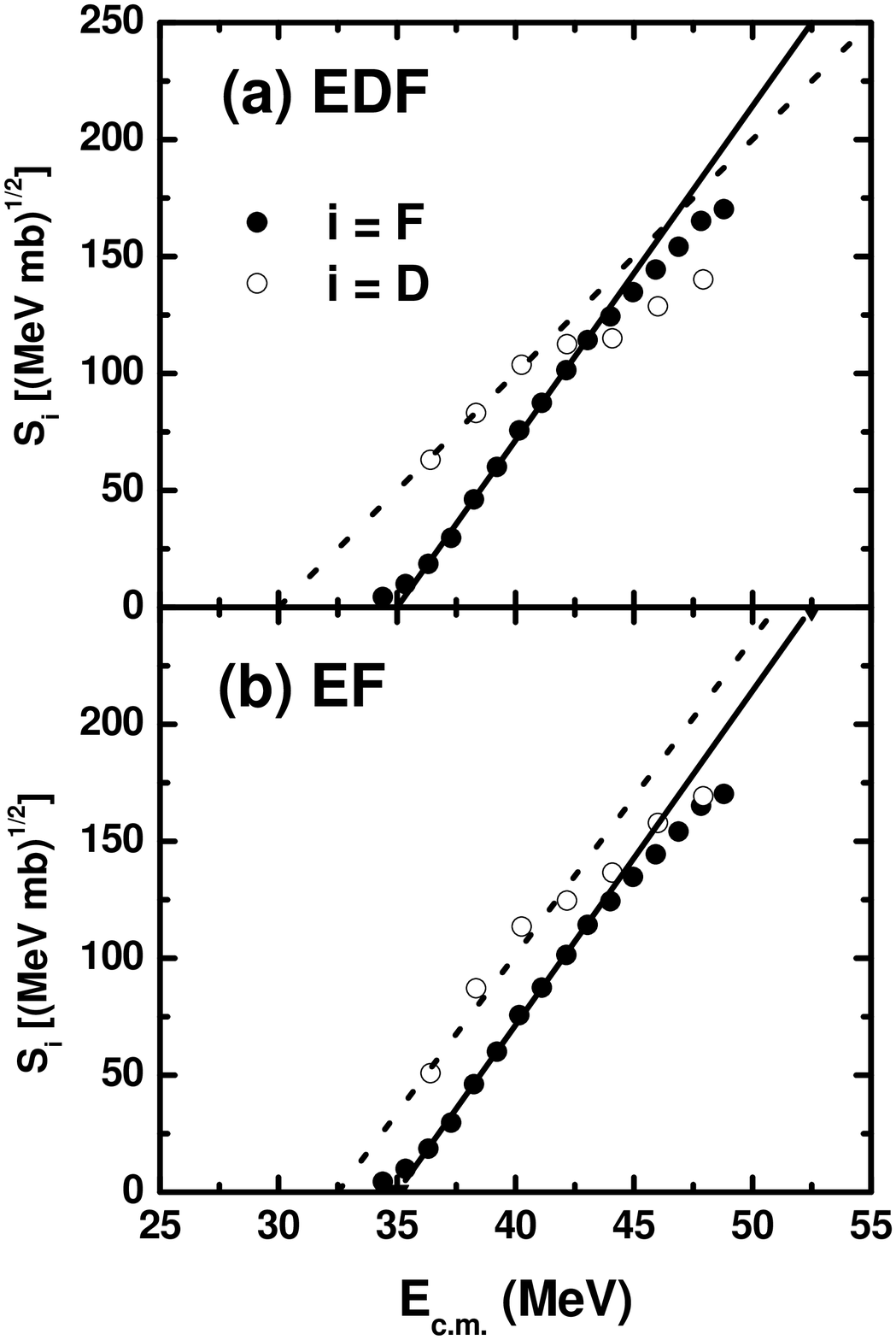}
\end{center}
\caption{\label{fig1} The Stelson plot of
$S_{i}=\sqrt{E_{c.m.}\sigma_{i}}$ for DR ($i=D$, open circles) and
fusion ($i=F$, solid circles) cross sections with (a) the
experimental and (b) the semi-experimental DR cross sections. The
straight lines are drawn to show the extraction of the threshold
energies $E_{0,i}$.}
\end{figure}


\begin{figure}
\begin{center}
\includegraphics[width=0.65\linewidth]{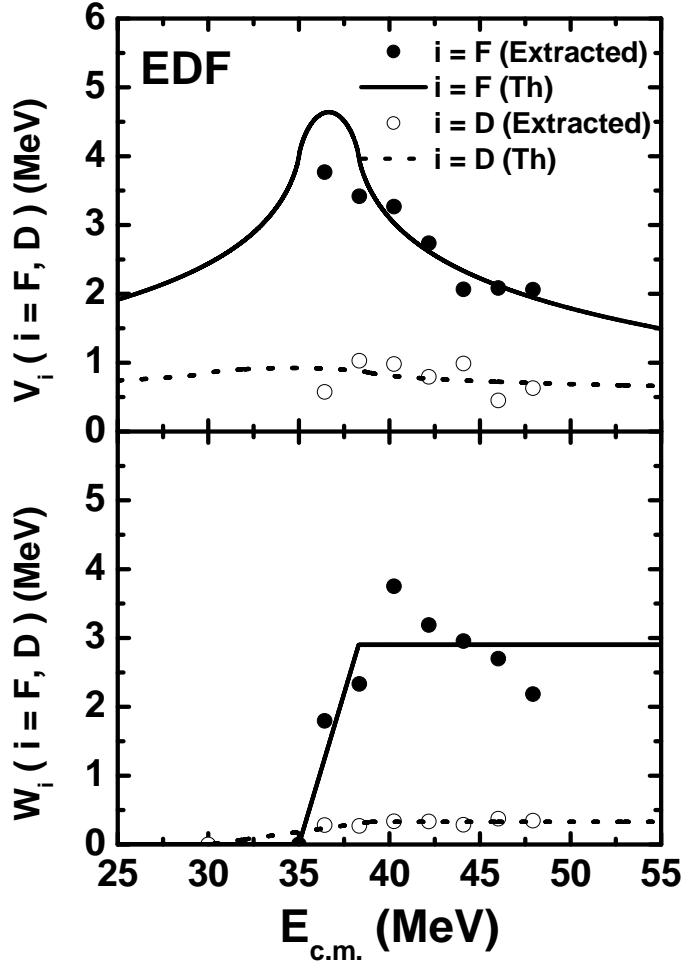}
\end{center}
\caption{\label{fig2}~The strength parameters $V_{i}$ (upper
panel) and $W_{i}$ (lower panel) for $i=D$ and $F$ as functions of
$E_{c.m.}$ in the EDF case. The open and solid circles are the
strength parameters for $i=D$ and $F$, respectively. The dotted
and solid lines in the lower panel denote $W_{D}$ and $W_{F}$ from
Eqs. (9) and (11), respectively, while the dotted and solid curves
in the upper panel represent $V_{D}$ and $V_{F}$ calculated by
using the dispersion relation of Eq. (6) with $W_{i}$ given by
Eqs. (9) and (11). The reference energies, $V_{F}(E_{s})$ and
$V_{D}(E_{s})$, are chosen as 4.0~MeV and 0.85~MeV, respectively.}
\end{figure}

\begin{figure}
\begin{center}
\includegraphics[width=0.65\linewidth]{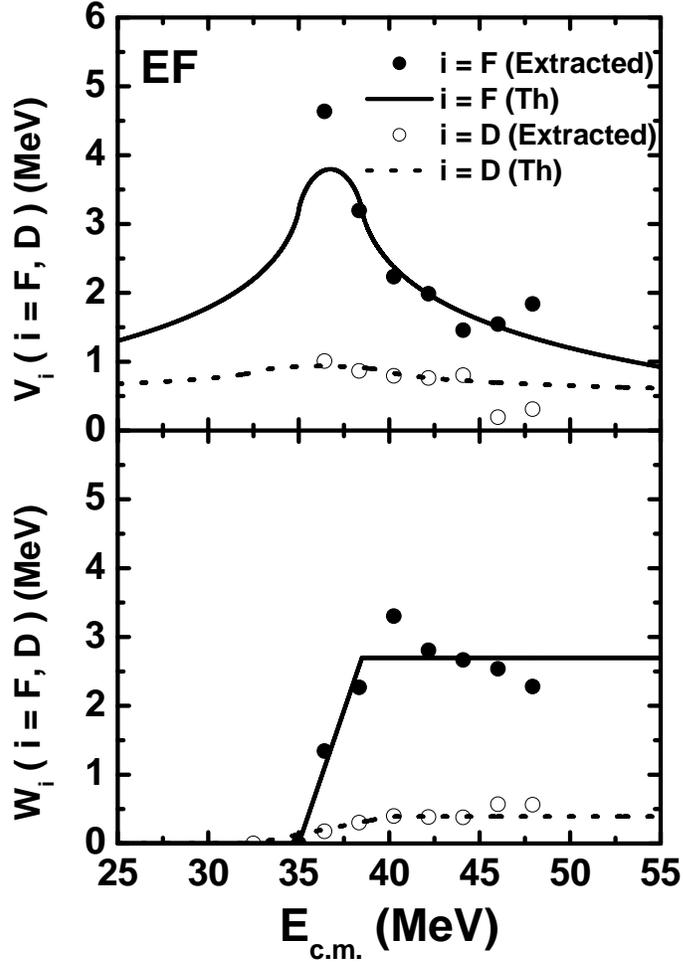}
\end{center}
\caption{\label{fig3}~The same as in Fig.~2, but for the EF case. 
The dotted (solid) line in the lower panel denotes $W_{D}$ ($W_{F}$) 
from Eq.~(10) (Eq.~(12)). The dotted (solid) curve in the upper panel
represents $V_{D}$ ($V_{F}$) obtained by the dispersion relation.
The reference energy $E_s$ for $V_{F}(E_{s})$ and $V_{D}(E_{s})$, 
are taken as 3.2~MeV and 0.85~MeV, respectively.}
\end{figure}

\begin{figure}
\begin{center}
\includegraphics[width=0.75\linewidth]{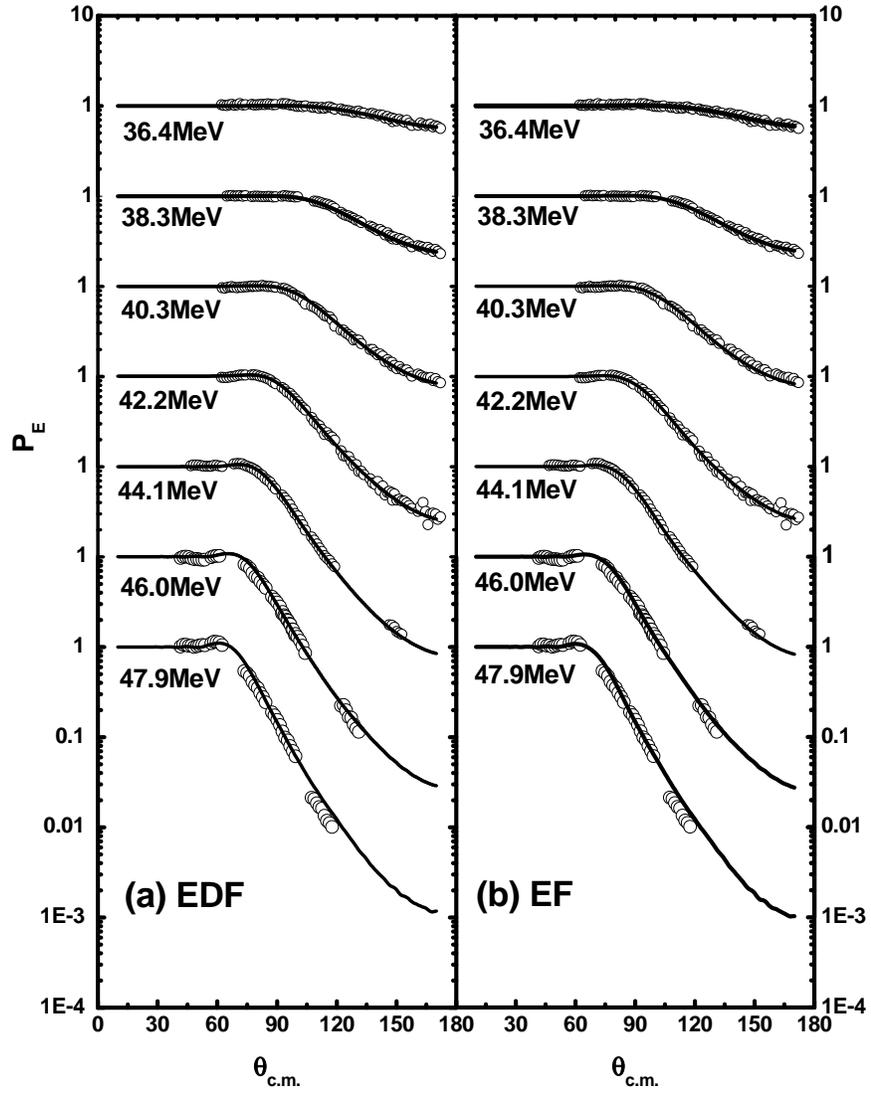}
\end{center}
\caption{\label{fig4}~Ratios of the elastic scattering cross
sections to the Rutherford cross section calculated with our final
dispersive optical potential for (a) the EDF and (b) the EF cases
are shown in comparison with the experimental data. The data are
taken from Ref.~\cite{woo1}.}
\end{figure}

\begin{figure}
\begin{center}
\includegraphics[width=0.70\linewidth]{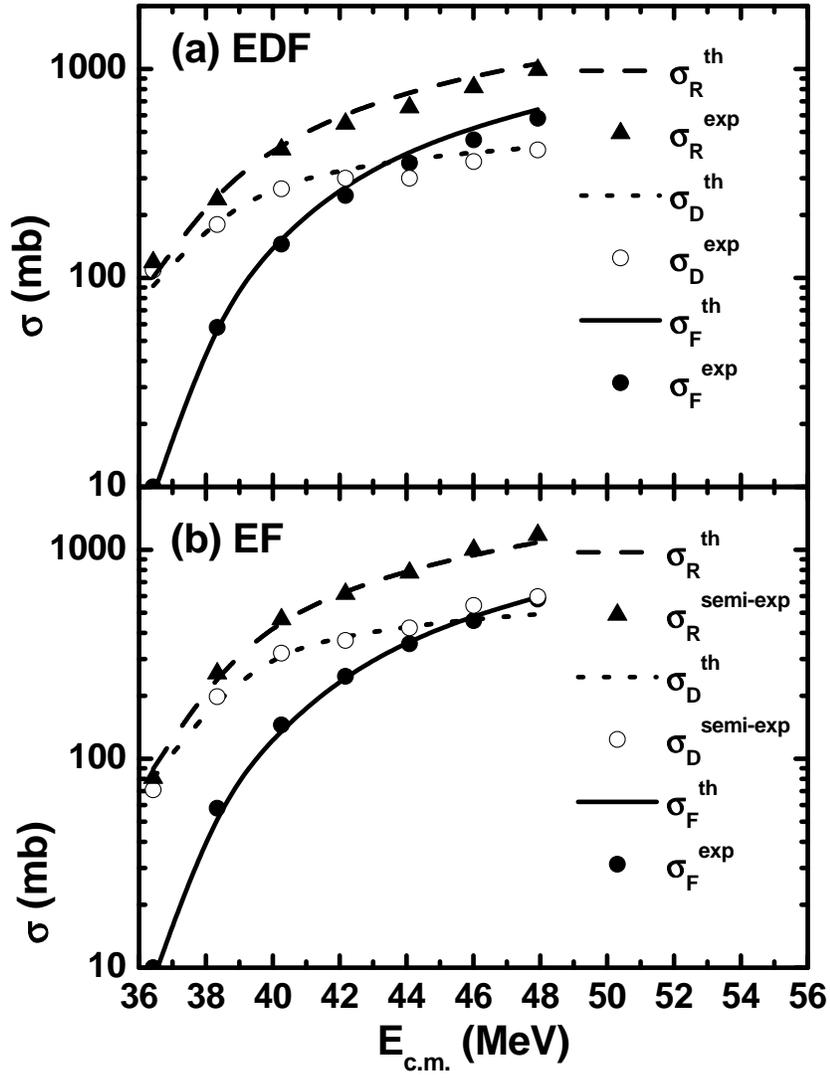}
\end{center}
\caption{\label{fig5} ~DR and fusion cross sections calculated
with our final dispersive optical potential for the (a) EDF and
(b) EF cases are shown in comparison with the experimental data.
$\sigma_{D}^{semi-exp}$ denoted by the open circles in EF case are
obtained as described in Sec.II. $\sigma_{D}^{exp}$ denoted by the
open circles in EDF case are the experimental DR cross
sections~\cite{woo2}. The fusion data are from Ref.~\cite{das2}.}
\end{figure}

\newpage

\setlength{\leftmargin}{6em} \vspace*{4.0cm}
\begin {center}
FIGURE CAPTIONS
\end {center}

\vspace {1ex}\hfill\break
\par
Fig.~1.~The Stelson plot of $S_{i}=\sqrt{E_{c.m.}\sigma_{i}}$ for
DR ($i=D$, open circles) and fusion ($i=F$, solid circles) cross
sections with (a) the experimental and (b) the semi-experimental
DR cross sections. The straight lines are drawn to show the
extraction of the threshold energies $E_{0,i}$.

\vspace {1ex}\hfill\break
\par
Fig.~2.~The strength parameters $V_{i}$ (upper panel) and $W_{i}$
(lower panel) for $i=D$ and $F$ as functions of $E_{c.m.}$ in the
EDF case. The open and solid circles are the strength parameters
extracted by $\chi^2$-fitting for $i=D$ and $F$, respectively. 
The dotted and solid lines in the
lower panel denote $W_{D}$ and $W_{F}$ from Eqs. (9) and (11),
respectively, while the dotted and solid curves in the upper panel
represent $V_{D}$ and $V_{F}$ calculated by using the dispersion
relation of Eq. (6) with $W_{i}$ given by Eqs. (9) and (11). 
The reference energy $E_s$ for $V_{F}(E_{s})$ and $V_{D}(E_{s})$, 
are chosen as 4.0~MeV and 0.85~MeV, respectively.

\vspace {1ex}\hfill\break
\par
Fig.~3.~The same as in Fig.~2, but for the EF case. The dotted
(solid) line in the lower panel denotes $W_{D}$ ($W_{F}$) from
Eq.~(10) (Eq.~(12)). The dotted (solid) curve in the upper panel
represents $V_{D}$ ($V_{F}$) obtained by the dispersion relation.
The reference energy $E_s$ for $V_{F}(E_{s})$ and $V_{D}(E_{s})$, 
are taken as 3.2~MeV and 0.85~MeV, respectively.

\vspace {1ex}\hfill\break
\par
Fig.~4.~Ratios of the elastic scattering cross sections to the
Rutherford cross section calculated with our final dispersive
optical potential for (a) the EDF and (b) the EF cases are shown
in comparison with the experimental data. The data are taken from
Ref.~\cite{woo1}.

\vspace {1ex}\hfill\break
\par
Fig.~5.~DR and fusion cross sections calculated with our final
dispersive optical potential for the (a) EDF and (b) EF cases are
shown in comparison with the experimental data.
$\sigma_{D}^{semi-exp}$ denoted by the open circles in EF case are
obtained as described in Sec.II. $\sigma_{D}^{exp}$ denoted by the
open circles in EDF case are the experimental DR cross
sections~\cite{woo2}. The fusion data are from Ref.~\cite{das2}.



\newpage

\begin{references}

\baselineskip=22pt

\bibitem{so1} W. Y. So, S. W. Hong, B. T. Kim, and T. Udagawa,
Phys. Rev. C {\bf 69}, 064606 (2004).
\bibitem{kee1} N. Keeley, S. J. Bennett, N. M. Clarke, B. R. Fulton,
G. Tungate, P. V. Drumm, M. A. Nagarajan, and J. S. Lilly, Nucl.
Phys. {\bf A571}, 326 (1994).
\bibitem{wu1} Y. W. Wu, Z. H. Liu, C. J. Lin, H. Q. Zhang, M. Ruan, F. Yang,
Z. C. Li, M. Trotta, and K. Hagino, Phys. Rev. C. {\bf 68}, 044605 (2003).
\bibitem{das1} M. Dasgupta {\it et al.}, Phys. Rev. C {\bf 66}, 041602(R)
(2002).
\bibitem{sig1} C. Signorini {\it et al.}, Phys. Rev. C {\bf 61}, 061603(R)
(2000).
\bibitem{sig2} C. Signorini {\it et al.}, Eur. Phys. J. A {\bf 5}, 7 (1999) and
private communications.
\bibitem{uda1} T. Udagawa, B. T. Kim, and T. Tamura, Phys. Rev. C
{\bf 32}, 124 (1985); T. Udagawa and T. Tamura, Phys. Rev. C {\bf
29}, 1922 (1984).
\bibitem{hong} S.-W. Hong, T. Udagawa, and T. Tamura, Nucl. Phys.
{\bf A491}, 492 (1989).
\bibitem{uda2} T. Udagawa, T. Tamura, and B. T. Kim, Phys. Rev. C
{\bf 39}, 1840 (1989); B. T. Kim, M. Naito, and T. Udagawa, Phys.
Lett. B {\bf 237}, 19 (1990).
\bibitem{mcc1} R. H. McCamis {\it et al.}, Can. J. Phys. {\bf 64}, 685
(1986).
\bibitem{eli1} T. Eliyakut-Roshko, R. H. McCamis, W. T. H. van Oers,
R. F. Carlson, and A. J. Cox, Phys. Rev. C {\bf 51}, 1295 (1995).
\bibitem{auc1} A. Auce {\it et al.}, Phys. Rev. C {\bf 53}, 2919 (1996).
\bibitem{sin1} P. Singh, A. Chatterjee, S. K. Gupta, and S. S. Kerekatte,
Phys. Rev. C. {\bf 43}, 1867 (1991).
\bibitem{kol1} J. J. Kolata {\it et al}.,
Phys. Rev. Lett. {\bf 81}, 4580 (1998).
\bibitem{maha} C. C. Mahaux, H. Ngo, and G. R. Satchler, Nucl. Phys.
{\bf A449}, 354 (1986); Nucl. Phys. {\bf A456}, 134 (1986).
\bibitem{naga} M. A. Nagarajan, C. C. Mahaux, and G. R. Satchler,
Phys. Rev. Lett. {\bf 54}, 1136 (1985).
\bibitem{bae1} A. Baeza, B. Bilwes, J. Diaz, and J. L. Ferrero,
Nucl. Phys. {\bf A419}, 412 (1984).
\bibitem{lil1} J. S. Lilley, B. R. Fulton, M. A. Nagarajan, I. J. Thompson,
and D. W. Banes, Phys. Lett. {\bf 151B}, 181 (1985).
\bibitem{ful1} B. R. Fulton, D. W. Banes, J. S. Lilley, M. A. Nagarajan,
and I. J. Thompson, Phys. Lett. {\bf 162B}, 55 (1985).
\bibitem{stel} P. H. Stelson, Phys. Lett. B {\bf 205}, 190 (1988);
P. H. Stelson, H. J. Kim, M. Beckerman, D. Shapira, and R. L.
Robinson, Phys. Rev. C {\bf 41}, 1584 (1990).
\bibitem{woo1} R. J. Woolliscroft {\it et al.}, Phys. Rev. C {\bf 69},
044612 (2004).
\bibitem{das2} M. Dasgupta {\it et al.}, Phys. Rev. Lett {\bf 82}, 1395
(1999).
\bibitem{woo2}R. J. Woolliscroft {\it et al.}, Phys. Rev. C {\bf 68}, 014611
(2003).
\bibitem{brog} R. A. Broglia and A. Winther, {\it Heavy Ion Reactions Lecture
Note Volume I: Elastic and Inelastic Reactions}, p165, (Benjamins,
London, 1981).
\bibitem{love} W. G. Love, T. Terasawa, and G. R. Satchler,
Nucl. Phys. A {\bf 291}, 183 (1977).
\bibitem{kim1} B. T. Kim, W. Y. So, S. W. Hong, and T. Udagawa,
Phys. Rev. C. {\bf 65}, 044607 (2002).
\bibitem{sat1} G. R. Satchler, {\it Introduction to Nuclear Reactions},
(Wiley, New York, 1980).
\bibitem{wo77} H. Wojciechowski, L. R. Medsker, and R. H. Davis, Phys. Rev. C
{\bf 16}, 1767 (1977).
\bibitem{sat2} G. R. Satchler and W. G. Love, Phys. Rep. {\bf 55}, 183 (1979).
\bibitem{huss} M. S. Hussein, Phys. Rev. C {\bf 30}, 1962 (1984).
\bibitem{kim2} B. T. Kim, W. Y. So, S. W. Hong, and T. Udagawa,
Phys. Rev. C. {\bf 65}, 044616 (2002).
\bibitem{sig3} C. Signorini {\it et al.}, Prog. Theor. Phys. {\bf 107},
1 (2002).
\end{references}
\end{document}